\documentclass[page-classic]{epl2} 

\usepackage{amsmath}
\usepackage{amsfonts}

\usepackage{amssymb}
\usepackage{latexsym}

\newcommand{\pr}{f}

\newcommand{\fun}{\mu}
\newcommand{\ch}{\mathrm{ch}}
\newcommand{\sh}{\mathrm{sh}}
\renewcommand{\revision}{}

\title{Duality between preferential attachment and static networks on hyperbolic spaces}
\shorttitle{Duality between preferential attachment and hyperbolic networks} 

\author{L. Ferretti\inst{1,2} \and M. Cortelezzi\inst{3} \and M. Mamino\inst{4}}
\shortauthor{L. Ferretti \etal}

\institute{            
\inst{1} Syst\'ematique, Adaptation et Evolution (UMR 7138), UPMC Univ Paris 06, CNRS, MNHN, IRD, Paris, France \\
\inst{2} CIRB, Coll\`ege de France, Paris, France \\
\inst{3} Dipartimento di Fisica, Universit\`a di Pisa, Italy \\
\inst{4} CMAF, Universidade de Lisboa, 1649-003 Lisboa, Portugal
}

\abstract{There is a complex relation between the mechanism of preferential attachment, scale-free degree distributions and hyperbolicity in complex networks. In fact, both preferential attachment and hidden hyperbolic spaces often generate scale-free networks. 
We show that there is actually a duality between a class of growing spatial networks based on preferential attachment on the sphere and a class of static random networks on the hyperbolic plane. Both classes of networks have the same scale-free degree distribution as the Barabasi-Albert model. 
As a limit of this correspondence, the Barabasi-Albert model is equivalent to a static random network on an hyperbolic space with infinite curvature. }

\begin{document}

\maketitle

\section{Introduction}

Scale-free networks have attracted a lot of interest as models for many technological, social and biological networks. Their degree distribution has been explained by the mechanism of preferential attachment in growing networks \cite{barabasi1999emergence}. 
The same distribution can be obtained in models with a preferential attachment mechanism that includes other features of the nodes \cite{ferretti2012features}, for example their position in some physical or abstract space \cite{ferretti2011preferential,barthelemy2003crossover,manna2002modulated,xulvi2002evolving,yook2002modeling,flaxman2006,flaxman2007,jordan2010}.

On the other hand, static models based purely on hidden variables have also been proposed for scale-free networks, see for example  \cite{caldarelli2002scale,boguna2003class}. Other models generate an effective preferential attachment mechanism through optimization or competition, e.g. \cite{fabrikant2002,berger2004,d2007emergence}. 

Networks on hyperbolic spaces have recently attracted some interest as models for real-world scale-free networks. It was shown in \cite{krioukov2009curvature} that the degree distribution of static random networks on the hyperbolic plane $\mathbb{H}^2$ is naturally power-law. More general models based on static networks on hidden hyperbolic spaces \cite{serrano2008self,krioukov2008efficient,krioukov2009curvature} 
seem to be able to explain in a simple way the 
scale-free
degree distribution, assortativity and clustering observed for Internet and other real networks. 

It has often been suggested that there is a relation between the scale-free structure of a network and its property of negative curvature, or hyperbolicity. Several forms for this relation have been presented, some involving the network topology \cite{jonckheere2007upper,jonckheere2008scaled}, other building scale-free networks on hyperbolic spaces \cite{serrano2008self,krioukov2008efficient,krioukov2009curvature}. However, the role of preferential attachment has not been studied in details, 
with the exception of 
a recent work that shows how an effective preferential attachment mechanism emerges from the underlying hyperbolics space \cite{papadopoulos2012popularity}. In this paper, we take the reverse route and find dual hyperbolic models for a class of networks built on preferential attachment. More precisely, there is an equivalence between a class of growing networks with preferential attachment on hidden spheres $S^{D}$ 
and a class of static random networks on hyperbolic spaces $\mathbb{H}^{D+1}$. 
The equivalence is valid in the limit of large networks and it works only approximately for young nodes.
Moreover, as an interesting limit of this equivalence, we find a correspondence between the original Barabasi-Albert model and some random static network models on $\mathbb{H}^D$ in the strong curvature limit. Our results show that preferential attachment is deeply related to the hyperbolic structure of the abstract space behind the network.

\section{Networks with preferential attachment as hyperbolic networks}

We discuss the relation between 
two models: (PA) networks grown by preferential attachment on the $D$-dimensional sphere, and (SH) static random networks on the $(D+1)$-dimensional hyperbolic space. The PA models have already been studied in \cite{ferretti2011preferential,jordan2010}, while the SH models are modifications of the hyperbolic models proposed by Krioukov, Bogu\~n\'a and collaborators \cite{krioukov2009curvature,krioukov2010hyperbolic}. 
We relate them by matching both their ``node density'' and their ``connection probability'' in the proper space, thus showing the duality. The birth time of the nodes in the growing network is mapped to the radial coordinate of the hyperbolic space. 

\revision{The structure of the two models is summarized here:}
\begin{center}
\revision{\vspace{0.1cm}
\hspace{-1.6cm}
\begin{tabular}{|c|c|}
\hline
\emph{Preferential attachment:} & \emph{Static hyperbolic network:} \\
\hline
Parameters: $D$, $m$, $T$, $\gamma$  & $D$, $\zeta$, $\langle k \rangle$, $N$, $R$, $\alpha$\\
\hline
Start from small network &  \\
$\downarrow$ &  Choose $N$ positions at random\\ 
Choose a random position $x$ & according to the density (\ref{hdens}) \\
on the sphere $S^D$ & on the hyperbolic disk $H^D$ \\
$\downarrow$ & of curvature $\zeta$ and radius $R$  \\
Add a new node with $m$ &  $\downarrow$ \\
links in position $x$ & Add $N$ nodes  \\
$\downarrow$ &  in these positions\\
Connect the loose ends & $\downarrow$ \\
of the $m$ links  &  For each pair of nodes, \\
to random nodes & connect them  \\
chosen according to  & with the distance-\\
preferential attachment  & dependent  probability (\ref{connprobh}) \\
in eqs. (\ref{probconn}), (\ref{fd}) & multiplied by $\mu(N,R)$\\
$\downarrow$ & \\
Repeat until $T$ nodes & \\
\hline
\end{tabular} 
\vspace{0.1cm}
}
\end{center}

Note that all these networks are spatial networks. We will not focus on the spatial aspect here; the reader is referred to \cite{barthelemy2011} for an extensive review of the subject. 


\subsection{(PA) Networks with preferential attachment on $S^D$}

We consider a growing network on $S^{D}$ with a generic connection probability $\pr(\vartheta)$ and preferential attachment. At every time, a node $i$ is added to the system with $m$ links to $m$ nodes that are randomly chosen according to the probability
\begin{equation}
p_j=\frac{\pr(\vartheta_{i,j})k_j}{\sum_l\pr(\vartheta_{i,l})k_l}\label{probconn}
\end{equation} 
where $\vartheta_{i,j}$ is the angular distance between $i$ and $j$. If $\pr(\vartheta)$ is an integrable function on the sphere, then these models have the same degree distribution as the usual Barabasi-Albert model, i.e. $p(k)\sim k^{-3}$ 
\cite{ferretti2011preferential} and no clustering in the thermodynamic limit \cite{ferretti2012features}. We are particularly interested in the case of a connection function of the form
\begin{equation}
\pr(\vartheta)=(1-\cos(\vartheta))^{-\gamma}\label{fd}
\end{equation}
which is integrable for $\gamma<D/2$. We denote the solid angle (i.e. the position on the sphere) by $\Omega$.

The aim of this section is to show that this subclass of models is equivalent to a class of static models on hyperbolic space. We employ a trick reminescent of \cite{boguna2003class}, that is, we consider the birth time $t_0$ of each node as a spatial variable with uniform distribution between $1<t_0<T$. We can substitute the degree $k(t|t_0)$ with its meanfield average $k(t|t_0)=m\left(\frac{t}{t_0}\right)^{1/2}$ and then calculate the ``random connection probability'' that is the probability that an edge is formed between two nodes born at times $t_1<t_2$:
\begin{equation}
p(t_1,\Omega_1;t_2,\Omega_2)=\frac{\mathrm{Vol}(S^D) \pr(\vartheta_{1,2})}{\int d^D\Omega \pr(\vartheta)}m\frac{\left(\frac{t_2}{t_1}\right)^{1/2}}{\int_{t'<t_2} dt' \left(\frac{t_2}{t'}\right)^{1/2}} 
\end{equation}
with
\begin{equation}
\int_{t_1<t_2} p(t_1,\Omega_1;t_2,\Omega_2)dt_1\frac{d^D\Omega_1}{\mathrm{Vol}(S^D)}dt_2\frac{d^D\Omega_2}{\mathrm{Vol}(S^D)}=mT
\end{equation} 
and the connection probability in the space $(t_0, \Omega)$ is
\begin{align}
p(t_1,\Omega_1;&t_2,\Omega_2)=\frac{\mathrm{Vol}(S^D)m}{2\int d^D\Omega \pr(\vartheta)}\cdot\frac{\pr(\vartheta_{1,2})}{\left({t_1}{t_2}\right)^{1/2}}\\
\propto & \exp{\left[-\frac{1}{2}\left(\log(t_1)+\log(t_2)-2\log(\pr(\vartheta_{1,2}))\right)\right]}\nonumber
\end{align}
This estimate is reasonable if the node degrees are well approximated by their mean values, i.e. there is an error related to the variance of $k(t)$ \cite{krioukov2013}. The relative error on $t_0$ scales like $\sqrt{k(t|t_0)}|dt_0/dk|/t_0\sim k^{-1/2}$, which is small for nodes not too young and in the large network limit.   

Now we can map the space $(t_0, \Omega)$ to the hyperbolic space $\mathbb{H}^{D+1}$ and in particular to a disk of radius $R$. Consider the spherical coordinates $(r,\Omega)$ in hyperbolic space. The mapping is $(t_0,\Omega)\rightarrow (r,\Omega)$ with $t_0=\nu(R) e^{D\alpha r}$. The corresponding density in the hyperbolic space is
\begin{equation}
\rho(r)\,dr\,d^D\Omega=\frac{\nu(R) D\alpha\ e^{D\alpha r}}{\mathrm{Vol}(S^D)}\,dr\,d^D\Omega \label{denspa}
\end{equation} and the total number of nodes is $N=\nu(R) (e^{D\alpha R}-1)=T$. The ``random connection probability'' has the form 
\begin{align}
p(r_1,\Omega_1;&r_2,\Omega_2)=\frac{\mathrm{Vol}(S^D) m}{2\nu(R) \int d^D\Omega (1-\cos(\vartheta))^{-\gamma}}\cdot\label{connprobpa}\\ \cdot \exp&\left[{-\frac{D \alpha}{2}\left(r_1+r_2+\frac{2\gamma}{D \alpha}\log(1-\cos(\vartheta_{1,2}))\right)}\right]
\nonumber \end{align}
In this mapping, the function $\nu(R)$ and the parameter $\alpha$ can be chosen arbitrarily. 

In the next section we will see that the density (\ref{denspa}) and the connection probability (\ref{connprobpa}) correspond precisely to the ones of a class of random network models in $\mathbb{H}^{D+1}$. 

\subsection{(SH) Static networks on hyperbolic spaces}

We present a class of static random networks that is a 
variant of the $\mathbb{H}^2$ model in \cite{krioukov2009curvature}. We consider a network with $N\gg1$ nodes uniformly distributed on a hyperbolic space $\mathbb{H}^D$ of constant negative curvature $K=-\zeta^2$. The nodes occupy random positions in a ball of radius $R\gg 1/\zeta$. This space can be described by a radial coordinate $r$ and $D-1$ angular coordinates $\Omega$ describing the position on the sphere $S^{D-1}$ of constant radius. The volume density is $J(r)\propto\sh(\zeta r)^{D-1}$ where $\sh(x)$ denotes the hyperbolic sine of $x$, and for large $r$ it can be approximated as $
J(r)\simeq e^{(D-1)\zeta r}$. We allow for a nonhomogeneous distribution of nodes, in particular we consider the density of nodes increasing exponentially with the radius
\begin{equation}
\rho(r)\,dr\,d^{D-1}\Omega\simeq\frac{N(D-1)\alpha\, e^{(D-1)\alpha r}}{\mathrm{Vol}(S^{D-1})\left(e^{(D-1)\alpha R}-1\right)}\,dr\,d^{D-1}\Omega\label{hdens}
\end{equation} 
The distance between two points $(r_1,\Omega_1)$ and $(r_2,\Omega_2)$ is given by
\begin{align}
\ch(\zeta d_{1,2})=\ch(\zeta r_1)\ch(\zeta r_2) -\sh(\zeta r_1)\sh(\zeta r_2)\cos(\vartheta_{1,2}) \label{hdist}
\end{align}
where $\vartheta_{1,2}$ is the angular distance between the points $\Omega_1$ and $\Omega_2$. 

We build the network by randomly connecting pairs of nodes at distance $d$ with probability $p=\fun(N,R)p(d,R)$. The factor $\fun(N,R)$ gives the freedom to tune the average degree $\langle k \rangle$ to any chosen value, in our case $\langle k \rangle=2m$. A class of $D=2$ models with $p(d,R)$ a integrable function of $\chi=e^{(D-1)\zeta(d-R)/2}$ on $\mathbb{R}^+$  are equivalent to the $S^1$ models of \cite{serrano2008self} and have scale-free degree distributions \cite{krioukov2009curvature}. 
Here we consider a different subclass of models with a density of nodes dependent on $N$ as $N\simeq \nu(R) e^{(D-1)\alpha R}$, a value of $\alpha<\zeta$ and a random connection probability
\begin{equation}
p(d,R)=\frac{
1}{\nu(R)} e^{-(D-1)\alpha d/2}\label{connprobh}
\end{equation}
Choosing $\nu(R)=e^{-(D-1)\alpha R/2}$, these models have the same form as the models in \cite{krioukov2009curvature} but with $p(d,R)\propto e^{-(D-1)\alpha (d-R)/2}=\chi^{-\alpha/\zeta}$ that is a non-integrable function. From a direct calculation we find that the average degree for nodes at radius $k$ is $\bar{k}(r)\simeq e^{-(D-1)\alpha r/2}$, therefore also this model has a scale-free degree distribution $p(k)\simeq\rho(r(k))\left|\frac{dr(k)}{dk}\right|\propto k^{-3}$. 

Now we compare these random network models on the $(D+1)$-dimensional hyperbolic space and the growing network models on the $D$-dimensional sphere discussed in the previous section. It is clear that the density (\ref{hdens}) is identical to the density (\ref{denspa}) of the growing model. For the connection probability, a good approximation to the distance (\ref{hdist}) for $\vartheta \gtrsim e^{-\zeta\min(r_1,r_2)}$ is given by the formula
\begin{equation}
d=r_1+r_2+\frac{1}{\zeta}\log(1-\cos(\vartheta)) \label{approxdist}
\end{equation}
up to an irrelevant constant $-\log(2)/\zeta$, and substituting it in the expressions (\ref{connprobh}) we obtain exactly the random connection probability (\ref{connprobpa}) with the condition 
\begin{equation}
\gamma=\frac{D\alpha}{2\zeta}
\end{equation} 
on the parameters of the models. 
For large networks, neglecting a infinitesimal fraction of central nodes, the fraction of neighbours with $\vartheta < e^{-\zeta\min(r_1,r_2)}$ becomes arbitrarily small and so the fraction of incorrect links too. 
This proves the equivalence of the two models.

As a check of the correspondence, we note that the average degree for nodes depends on the birth time/position in the same way in both models, i.e. $e^{-D\alpha r/2}\sim t_0^{-1/2}$.

Moreover, all hyperbolic models in this class with the same dimension $D$ and the same ratio $\alpha/\zeta$ are equivalent in the large $N$ limit, because they are all equivalent to the model with preferential attachment on the sphere $S^D$ and $\gamma=(D/2)\cdot\alpha/\zeta$. In the hyperbolic models, the equivalence is just a rescaling of the metric. \revision{Note also that hyperbolic models with the same $N$ but different radii $R$ and different $\nu(R)$ are equivalent.} 

Since the preferential attachment mechanism does not produce a finite clustering in the limit of large size \cite{ferretti2012features}, the same is true for these dual hyperbolic networks, as can be verified directly \cite{marianprivate}. The null clustering can be seen as a consequence of the non-integrability of $p(d,R)\propto \chi^{-\alpha/\zeta}$ in the context of the $S^1$ models in \cite{serrano2008self,krioukov2009curvature}.

\revision{\subsection{Numerical results}

The duality presented above is valid for large networks under some approximations. 
We can
estimate the goodness of the approximations by comparing different quantities on the two sides of the duality. We simulate the PA model on the circle $S^1$ and the SH model on the hyperbolic disk $H^2$, which are dual to each other. We fix $m=5$ (i.e. $\langle k \rangle=10$) and we choose arbitrarily $\nu(R)=1$ and $\zeta=1$. The tunable parameter of the matching is therefore $\gamma=\alpha/2$.

A first check involves the quantity $\mu(N,R)$ in the SH model, which can be estimated analytically from the PA model by comparing eqs. (\ref{connprobpa}) and (\ref{connprobh}), obtaining 
\begin{equation}
\mu(N,R)=\frac{\langle k \rangle \mathrm{Vol}(S^D)}{2^{2+\alpha/2}\int d^D\Omega (1-\cos(\vartheta))^{-\gamma}}\label{mupred}
\end{equation}
independent on $N$ and $R$. This prediction is sensitive to the accuracy of the approximation (\ref{approxdist}). 
Numerical results for the deviations from this prediction are presented in Fig.~\ref{fig-pnorm}. The agreement is very good for large network size ($>10^5$), while for small networks and $\gamma$  close to $1/2$ there are strong finite-size effects.  
\begin{figure}
\onefigure[width=0.5\columnwidth]{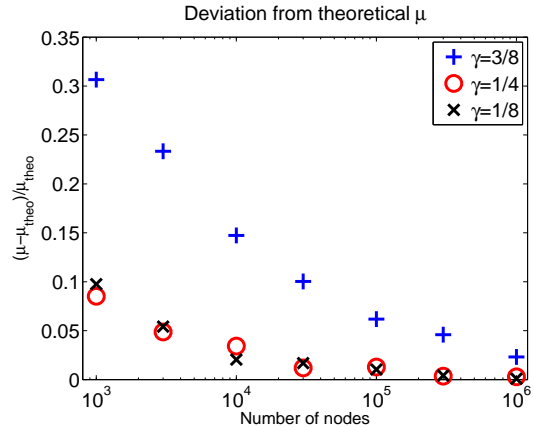}
\caption{Relative deviation of $\mu(N)$ from the prediction in eq.~(\ref{mupred}) in the SH model for different values of $\gamma$.}
\label{fig-pnorm}
\end{figure}

We compare three important features of these models across the duality: (i) the degree distribution, Fig.~\ref{figpk}; (ii) the assortativity in degree, or degree-degree correlation, expressed as the average degree of neighbours $\langle k_{NN}\rangle$ as a function of the node degree $k$, in Fig.~\ref{fig-a}; (iii) the clustering coefficient as a function of the size of the network in Fig.~\ref{fig-cl}. 

The degree distribution is roughly independent on $\gamma$ and matches very well across the duality, except for very small nodes. The assortativity is weak in these networks, which is reflected in the flatness of $\langle k_{NN}(k)\rangle$, but the general trend is captured by the duality - except for the disassortative tail at low $k$ appearing in PA models, which does not match the behaviour of the SH models. Instead, the latter have a flatter curve that agrees with the prediction for the PA model from continuum approximation \cite{ferretti2012features}. Finally, both the absolute value and the time scaling of the clustering coefficient match very well between the two sides of the duality. 
Note that the clustering matches well also for $\gamma>1/3$, where the theory in \cite{ferretti2012features} breaks down and the scaling is different from the Barabasi-Albert model due to cutoff-dependent effects.
\begin{figure}
\onefigure[width=0.5\columnwidth]{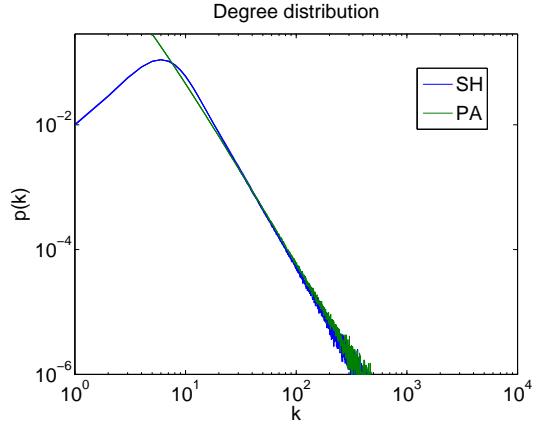}
\caption{Degree distribution for the two dual models at $N=10^6$. The different values of $\gamma=1/8$, $1/4$, $3/8$ result in the same distribution.}
\label{figpk}
\end{figure}
\begin{figure}
\onefigure[width=0.5\columnwidth]{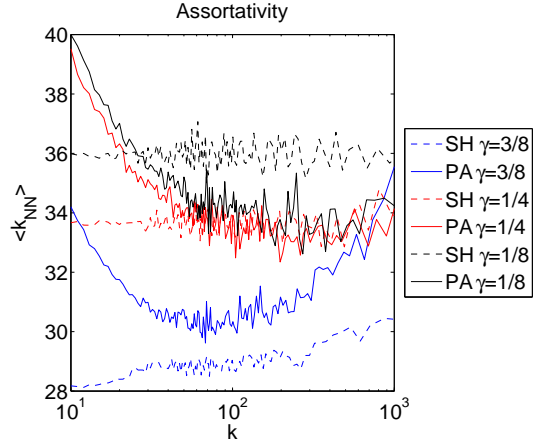}
\caption{Average neighbour degree $\langle k_{NN}\rangle$ for the two dual models, for $N=10^6$ and different values of $\gamma$.} 
\label{fig-a}
\end{figure}
\begin{figure}
\onefigure[width=0.5\columnwidth]{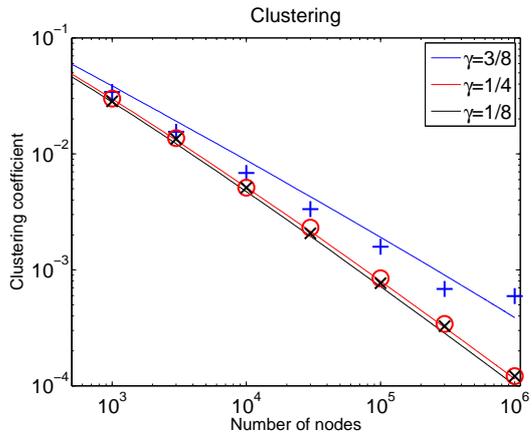}
\caption{Clustering coefficient for the two dual models - the PA model (continuous lines)  and the SH model (points) - as a function of the network size, for different values of $\gamma$.}
\label{fig-cl}
\end{figure}
}

\section{Barabasi-Albert model as network on 
infinitely curved hyperbolic space} 

There are two interesting limit of the above correspondence from the point of view of the growing network model. The first limit is obtained by sending $D\rightarrow\infty$ with constant $\gamma$: 
the growing model in this limit lives on an $S^\infty$ sphere, which has the curious property that the angular distance between two random points is always $\pi/2$, therefore the spatial structure disappears and the model reduces to the original Barabasi-Albert model. The second limit can be obtained by sending $\gamma\rightarrow 0$ for a given dimension $D$: in this case the connection function is $\pr(\vartheta)=1$, therefore we recover once again the usual Barabasi-Albert model.

On the other side, the corresponding models on $\mathbb{H}^D$ are in the strong curvature limit $\zeta/\alpha=D/\gamma\rightarrow\infty$ and have a very simple structure: given the product $\lambda=D\alpha$, they live on hyperbolic space with a metric structure given by the distance $d=r_1+r_2$, an exponential density of nodes $\rho(r)\sim e^{\lambda r}$ and  a connection probability $p\sim e^{-\lambda d/2}$. A simple calculation gives $\bar{k}(r)\sim e^{-\lambda r/2}$ and therefore $p(k)\propto k^{-3}$ also in this limit, as expected from the equivalence.

The Barabasi-Albert model of growing random networks is therefore equivalent to a class of static random network models on the finite or infinite-dimensional homogeneous hyperbolic space $\mathbb{H}^D$ or $\mathbb{H}^{\infty}$ in the infinite curvature limit.

\section{Conclusions}

The initial question for this work was to know if the networks generated by preferential attachment are someway equivalent to hyperbolic networks. 
We have shown that the answer is positive at least for some classes of networks, including the interesting case of the Barabasi-Albert model. 
These results strengthen the relation between preferential attachment and hyperbolic spaces.


It is interesting that a mapping exists between growing and static networks, despite the difference in the equilibrium vs non-equilibrium nature of these systems. Recently, it has been shown that similar (but exact) dualities between growing and static random network can be found in a variety of contexts \cite{krioukov2013}. These relations between equilibrium and non-equilibrium systems are not uncommon in networks, as shown by the classical example of Bose-Einstein condensation phase transition in growing networks out of equilibrium \cite{bianconi2001bose}, but they are nevertheless quite promising in revealing common laws and unexplained similarities between widely different mechanisms.

\revision{These exact or approximate dualities have several interesting consequences. A first, general implication of the existence of these dualities is the impossibility to discriminate between static and growing networks models from a single time point. In fact, even for networks that fit well a simple static model, there could be an equivalent growing model that could therefore explain equally well the data, and vice versa. Discriminating between static and growing networks requires information from multiple time points.

A second consequence is the possibility to translate directly the results obtained in a model to the dual one. We already showed that it is possible to compute analytically quantities like the average degree in the dual model. Relevant features like degree distribution, assortativity and clustering can be estimated from either side of the duality and applied to the other. 
The dynamics and the critical points of phase transitions in models like Ising \cite{bianconi2002mean} or Bose-Hubbard \cite{halu2012phase} on a Barabasi-Albert network will also  be the same as on the static dual hyperbolic network at large curvature, and similarly for the resilience and resistance to attacks \cite{albert2000error, crucitti2003efficiency}, the critical parameters for infection or communication models \cite{pastor2001epidemic,pastor2001epidemic2}, etc. We expect that some of these results could be only weakly dependent on the exact shape of the connection function (\ref{fd}) or connection probability (\ref{connprobh}), therefore the matching could be qualitatively valid also for a larger class of models.

It would be of great interest to extend this duality further to hyperbolic models with finite clustering coefficient and to variations of the Barabasi-Albert model. Models with preferential attachment in an expanding space are among the potential candidates. 

More generally, exploration of approximate dualities in other simple network models with tunable parameters could lead to further insights on the nature of complex networks and a renewed look at the space of network models, similarly to what happened in string theory two decades ago with the discovery of string dualities \cite{polchinski1998string}.

% ...
}

%
%
%
%
%
%
%
%
%
%
%

\acknowledgments
We thank M. Bogu\~na, G. Marmorini and G. Bianconi for useful comments and discussions. L.F. acknowledges support from  grant ANR-12-JSV7-0007, Agence Nationale de la Recherche (France).

\bibliographystyle{eplbib}
\bibliography{networks}

%
%
%
%

\end{document}